# Molecular simulation-derived features for machine learning predictions of metal glass forming ability

Benjamin T. Afflerbach[a], Lane Schultz[a], John H. Perepezko[a], Paul M. Voyles[a], Izabela Szlufarska[a], Dane Morgan[a]

[a]University of Wisconsin-Madison

Abstract:

We have developed models of metallic alloy glass forming ability based on newly computationally accessible features obtained from molecular dynamics simulations. Since the discovery of metallic glasses, there have been efforts to predict glass forming ability (GFA) for new alloys. Effective evaluations of GFA have been obtained but generally relied on knowledge of alloy characteristic temperatures like the glass transition, crystallization, and liquidus temperatures but are of limited utility because these features require synthesizing and characterizing the alloy of interest. More recently, machine learning approaches to predict GFA have employed more accessible model features such as the elemental properties of constituent elements. However, these more accessible features generally provide less predictive accuracy than their less accessible counterparts. In this work we showed that it is possible to increase the predictive value of GFA models by using input features obtained from molecular dynamics simulations. Such features require only relatively straightforward and scalable simulations, making them significantly easier and less expensive to obtain than experimental measurements. We generated a database of molecular dynamics critical cooling rates along with associated candidate features that are inspired from previous research on GFA. Out of the list of 9 proposed GFA features, we identify two as being the most important to performance through a LASSO model. Enthalpy of crystallization and icosahedral-like fraction at 100 K showed promise because they enable a significant improvement to model performance and because they are accessible to flexible ab initio quantum mechanical methods readily applicable to almost all systems. This advancement in computationally accessible features for machine learning predictions GFA will enable future models to more accurately predict new glass forming alloys.

1. Introduction and motivation:

Since the discovery of the first melt quenched metallic glass in 1960, there has been a continuous search for new glassy alloys [1]. This search has yielded glassy alloys of scientific and commercial interest [2–4]. While successful, the rate of discovery of new alloys has been slow and only a small fraction of the possible search space has been explored [5]. The search for new glassy alloys can be broadly classified into two main methodologies. The first employs qualitative predictions of glass forming ability (GFA) using a variety of criteria such as searching for deep eutectics and identifying alloys with large atomic size mismatches [6]. The second methodology is focused on quantitative predictions of GFA which is measured by two main metrics: the critical casting diameter ($D_C$) or the related critical cooling rate ($R_C$). This study focuses on the second category of quantitative models, which give direct predictions of a GFA metric.

Quantitative models for GFA have taken different forms during their development and can be compared based on two factors: the accessibility of features used as inputs for the model and the predictive value of the model. When looking at a typical materials discovery workflow, the accessibility of model inputs is important as it can define the size and scope of the search space that is available. The highest level of accessibility is for features of a target alloy that are known without requiring any significant computation or experiments. For example, recent machine learning based models that only rely on elemental properties of constituent elements use such highly accessible features [7–10]. At the lowest level of practical feature accessibility are properties that require synthesizing and experimentally characterizing properties of the target alloy in the glassy state (which we will simply call measured properties). Examples of these include various characteristic temperatures of a glass such as the glass transitions temperature ($T_g$), the crystallization temperature ($T_x$), and the liquidus temperature ($T_l$). While informative, these features are significantly harder to obtain and present challenges in a materials discovery workflow because of the necessity to synthesize a glass prior to evaluating its GFA. The predictive value of a model covers both the quality of predictions (domain and accuracy) of a model as well as what GFA metric a model is designed to predict. The two most direct and most common metrics of GFA are the critical casting diameter ($D_C$) and critical cooling rate ($R_C$). Both offer high value as they are direct measures of GFA. Another measure of GFA with lower predictive value is the ability of an alloy to form a glass under specific cooling conditions. For example, recent machine learning models to predict

GFA have predicted a probability of forming a glass under melt spinning conditions, which can be very useful, but is not as direct a measure of GFA as a prediction of $D_C$ or $R_C$ [7].

Across this spectrum of input features and predictive value most models fall into two camps. They either use easily accessible features and sacrifice predictive value or they use features that are relatively inaccessible and can maximize predictive value. If more accessible features were available that maintained their information related to GFA, it would potentially be possible to generate machine learning models that have the best of both camps, *i.e.*, easy predictions that are highly accurate. The spectrum of models along these feature accessibility and prediction value metrics is summarized in Figure 1, and the goal of this work is to help develop models that are in the upper right quadrant. Models that fall in the bottom left quadrant would generally be uninteresting and not useful, and current models fall either in the top left or bottom right quadrants.

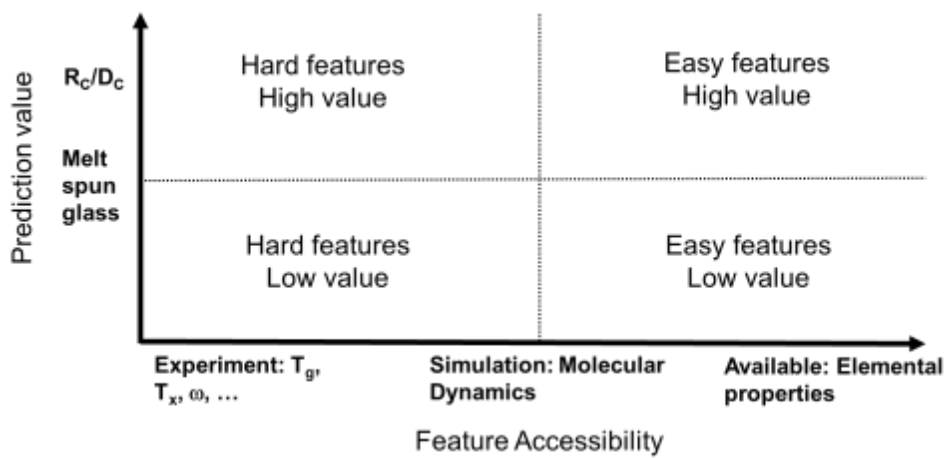

Figure 1. Outline of Feature Accessibility and Predictive Value Spectrum.

To find a middle ground in which predictive value is maintained while increasing feature accessibility, we used high throughput computational simulations to generate a range of features. These features are all generated by performing Molecular Dynamics (MD) simulations to quench a variety of metal alloys to extract information from the quench run or the resulting amorphous structure. In an ideal world, we would be able to perform these simulations on alloys with experimentally measured GFA metrics to use as the training and testing data. However, the overlap between alloys with measured GFA metrics and available interatomic potentials is relatively small. Therefore, we calculated the $R_C$ values directly from MD simulations by performing a series of quenches at varying cooling rates to obtain the GFA data for training and testing. This approach comes with a few limitations. Due to simulation time limitations, the accessible cooling rates are $10^9$ K/s and above. This high value means that conclusions obtained here will be limited to low GFA alloys. However, the hope is that lessons learned in this range may give insight into GFA for higher GFA alloys as well. The second limitation is that available interatomic potentials are not necessarily fit to amorphous structures and may misrepresent them in some way. Because all the data calculated will be using the same types of potentials, the $R_C$ values and features are at least fully self-consistent, although they may not match experimental values.

2. Computational methods and database details:

A computational critical cooling rate database was generated using the large-scale atomic/molecular massively parallel simulator (LAMMPS) molecular dynamics software found at http://lammps.sandia.gov [11]. The critical cooling rates were estimated by performing temperature ramps from 2500K to 100K in 50K steps. The hold time at each step was varied to achieve cooling rates ranging from $10^{14}$ K/s to $10^{11}$ K/s. Eleven binary systems were investigated: Al-Ag, Al-Cu, Al-La, Al-Sm, Al-Zr, Cu-Mg, Cu-Ni, Cu-Zr, Mg-Y, Ni-Zr, and Pd-Si. These binaries were chosen from available binary EAM potentials due to overlap with existing known glass forming systems as well as to maximize variety of elements included[12]. For each binary system, compositions near the edges of the binary were simulated with a maximum of 10% of the minority element. This was done because we can only calculate $R_C$ if it is within the cooling rate range given above, which corresponds to what are generally considered bad glass formers. Bad glass formers are expected to occur near pure elements. For each composition and cooling rate, 10 cooling runs were performed to account for the stochastic nature of crystallization during cooling. To find $R_C$ we used the following approach. First, we identified the slowest cooling rate at which more than 50 percent of cooling runs amorphized, denoted $R_C$ (2), The cooling rate on sampling step slower than $R_C$ (2) was denoted $R_C$ (1). We then assumed that the fraction of runs that yielded amorphous material was a linear function of $R_C$ between $R_C$ (1) and $R_C$ (2) and solved for the cooling rate that gave exactly 50% of runs amorphizing. This was taken as $R_C$. From the compositions and cooling rates simulated, 78 critical cooling rates were found and used in a critical cooling rate database.

Ten highly accessible features are considered as inputs for machine learning models. The first nine (excluding the ELEM features discussed more below) are from simulations or could be obtained from simulations. These nine features will be referred to as the "GFA features" or "GFA feature set" going forward. The features are all inspired from previous research in modeling GFA. Some are directly taken from previous literature and some are modified to fit into a high-throughput simulation workflow. A summary of the GFA and ELEM features can be found in *Table 1*.

Table 1. Summary of GFA features generated for machine learning models.

| Feature | Source |
|---|---|
| 1. glass transition temperature ($T_g$) | cooling |
| 2. liquidus temperature ($T_l$) | phase diagram |
| 3. reduced glass transition temperature ($T_{rg}$) | previous features 1 and 2 |
| 4. atomic packing density (APD at 100K) | cooling |
| 5. icosahedral-like (ICO-like) fraction (at 1.2 $T_g$) | 1.2 $T_g$ hold |
| 6. icosahedral-like (ICO-like) fraction (at 100K) | cooling |
| 7. diffusivity (at 1.2 $T_g$ [D]) | 1.2 $T_g$ hold |
| 8. variance of voronoi Polyhedra (Var) | cooling |
| 9. enthalpy of crystallization | energy minimization |
| 10. elemental features (ELEM) | MAST-ML |

We obtained the glass transition temperature $T_g$ from the $10^{14}$ K/s rate cooling run for all systems. $T_g$ values obtained at this high cooling rate are known to have a significant shift from experimental glass transition temperatures which may limit usefulness of this computational $T_g$, but there may be useful information that can be extracted. Furthermore, getting an estimate of $T_g$ is necessary for several of the other features which are defined in relation to $T_g$.

The liquidus temperature ($T_l$) is the second feature that has been used in some of the earliest models to predict GFA[13]. This feature was extracted from existing phase diagrams and not directly computed. While obtaining values in this way requires experiments at some point, they can be extrapolated from existing data so effectively that we consider them to be highly accessible. The liquidus values could also be simulated with MD with reasonable fidelity if needed.

The reduced glass transition temperature ($T_{rg} = T_g / T_l$) is one of the oldest features used to predict GFA [13]. Using this simple combination of the previous two features, we can potentially learn if this ratio improves learning compared to the base features that compose it. If a highly complex machine learning model was generated, we might expect this relationship to be learned. However, because a simple model is being used, providing the relationship directly may improve the model.

The atomic packing density (APD) feature has been shown to affect the mechanical properties of metallic glasses and has been loosely been tied to GFA [14–16]. The APD is calculated from the resulting amorphous structure after the 75 K/ps cooling runs and is averaged over the 10 runs. Similar to other features, the final structure was taken after a final energy minimization of that resulting structure. Metallic atomic radii and empirical atomic radii where used for the APD calculations are taken from the pymatgen materials analysis library [17].

The Icosahedral-like (ICO-like) fraction has been used as a feature linked to glass forming ability in a variety of systems [18–21]. In this work, we calculated two variations of ICO-like fraction. The first was obtained at 1.2 times the $T_g$ feature calculated previously. The second was taken from a snapshot of the final structure after cooling to 100K at the 75 K/ps cooling rate. In both cases, static energy minimization was performed in order to more clearly observe the underlying structure. In our definition of ICO-like fraction, we used the definition proposed by Bokas et al [15]. Icosahedral like clusters are identified based on Voronoi polyhedral indices of each atom and the ICO-like distinction is a slightly broader inclusion than the pure <0,0,12,0> indices. The fraction is the sum of atoms classified as ICO-like divided by the total atoms in the simulation.

The diffusivity (D) feature is the average self-diffusivity of the atoms in the system, regardless of composition. It was found by performing an additional temperature hold at 1.2 $T_g$. This feature is different than previous features because it is not trying to mimic a feature directly suggested by previous research. Instead, D is calculated as a way to include a kinetic feature whereas many of the others included are structural. A multiple time origins approach was used to calculate self-diffusion using the Einstein relation on mean squared displacement [22,23].

The Variance of Voronoi Polyhedra (Var) is a feature recently explored by Wang et al. [24] as a way to analyze the liquid structure and gain information about GFA. The variance of cluster fractions is defined as:

$$\sigma^2 = \frac{\sum_i (X_i - \mu)^2}{N_c}, \qquad (1)$$

where $X_i$ is the fractional contribution of a cluster type to the structure, $\mu$ is the average of the cluster fractions, and $N_c$ is the number of cluster types included in the calculation. As $N_c$ increases, cluster types with progressively lower fractions are included in the metric. This leads to a natural maximum in a plot of variance with respect to number of clusters included. In our work, this maximum variance is used as the variance metric as opposed to variance at a fixed number of clusters as was done by Wang et al. [24]. This modification was done because otherwise the metric varied widely from system to system with respect to the number of polyhedral averaged making it impossible to select a specific converged value for $N_c$.

The Enthalpy of Crystallization ($\Delta H_{crystal}$) is a feature inspired by related investigations of the competition between crystalline and amorphous phases [10,25,26]. In previous work, enthalpies have been obtained experimentally. Here we calculated a simple approximation of the enthalpy difference by taking the difference between the final amorphous structure obtained after the 75 K/ps cooling runs and 3 candidate crystal structures. $\Delta H_{crystal}$ is then defined as the minimum enthalpy difference between the amorphous structure and the three crystals. The three crystal structures generated are BCC, HCP, and FCC solid solutions of the relevant composition. Elements are randomly assigned to lattice positions in the appropriate ratios and the lattice volume is relaxed while fixing atoms on their lattice sites.

The Elemental Feature (ELEM) set is a set of elemental features compiled by Ward et al. and has been shown to be useful in a variety of applications in predicting materials properties purely from compositional information [9]. We used this feature set in combination with the MAterials Simulation Toolkit – Machine Learning (MAST-ML) [27,28] to generate a list of compositional features that includes the composition average, arithmetic average, minimum, maximum, and difference of the elemental features. Thus this feature entry in Table 1 actually corresponds to a long list of features that require no simulations to generate.

One common GFA criteria that was not included is the onset of crystallization temperature ($T_X$) [29–33]. This feature has not been simulated for a few different reasons which make it impractical to obtain in high-throughput fashion. The first is that $T_X$ is known to be highly sensitive to the heating rate. Previous experimental investigations such as those by Zhuang et al. have shown that changes in heating rate from 5 K/s to 80 K/s can result in a change of around 20K in $T_X$ [34]. The simulated heating rates, which if assumed to be on the same scale as the slowest cooling rates achieved in this work, would be eight orders of magnitude faster than those experiments and likely lead to large errors in $T_X$. In some cases one can extrapolate from computational time scales to experimental time scales through generating Time-Temperatture-Transformation (TTT) curves from isothermal holds at multiple temperatures, such as those completed by Louzguine-Luzgin, and Bazlov [35]. However this approach requires many simulations for each material, and the approach has only been demonstrated on pure elements Iron and Copper. For any alloys with better GFA it is expected that simulation time for these holds would increase dramatically. Because of these concerns $T_X$ was not pursued as a feature in this work. However, in the future it may be possible to add this to the feature set explored here.

All the models built used the MAST-ML machine learning toolkit which uses sci-kit learn implementations, models, and analysis tools [27,28]. The following model types of varying complexities were investigated: LASSO, Ridge Regression (RR), Random Forest (RF), and Gaussian Kernel Ridge Regression (GKRR). For each model, the hyperparameters were optimized via grid search and the performance was estimated with 5-fold cross validation. Using the GFA feature set, all models gave 5-Fold root mean squared errors within observed variance due to random splitting of train and test splits. Therefore, one model was chosen to report results on in this investigation. The LASSO model was chosen for its simplicity and ease of interpretability. Two models were built. The first model, referred to as the baseline model, uses only the ELEM set. The second model uses both ELEM and the range of GFA inspired features discussed above.

3. Results and Discussion:

The results from training on the ELEM features model are shown in Figure 2. The parity plot shown is the average result from 20 individual 5-Fold cross validation splits and the error bars are the standard deviation of predictions across those 20 splits. The lack of correlation between the predictions and true values makes it immediately clear that the model does not perform well and has essentially no predictive value beyond getting the average value. The Root Mean Squared Error (RMSE) divided by the standard deviation in the training set ($\sigma_y$) value approaches one, demonstrating that errors in predictions are on the same scale as the spread in the training data.

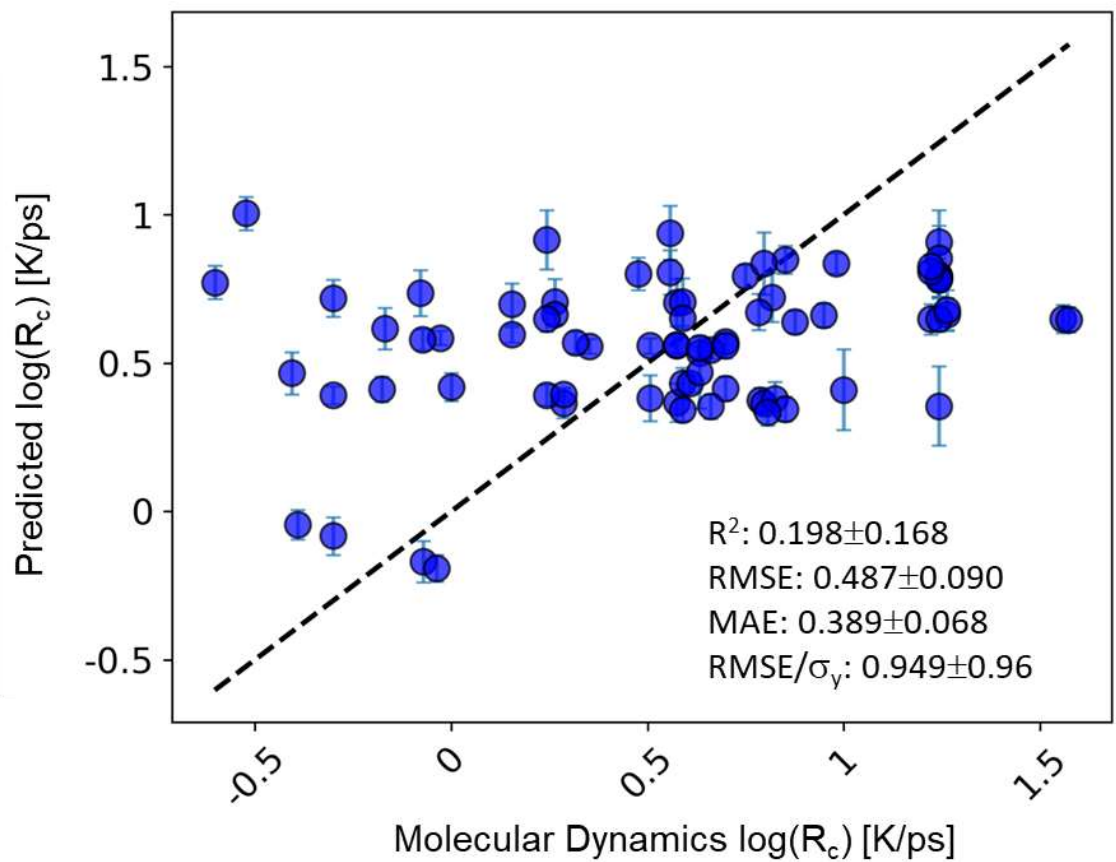

Figure 2. 5-Fold cross validation predictions of the baseline LASSO model. Error bars show the standard deviation in predictions over 20 cross validation runs. Error metrics are averages of individual statistics from each cross validation run.

Figure 3 shows the results of the model with the full feature set (ELEM and GFA features). Qualitatively, it is clear that the model has improved. The RMSE divided by $\sigma_y$ has dropped considerably when compared to the elemental-only model. Now, the RMSE is about half of the spread in the data. The $R^2$ value has also increased to 0.77 from 0.20.

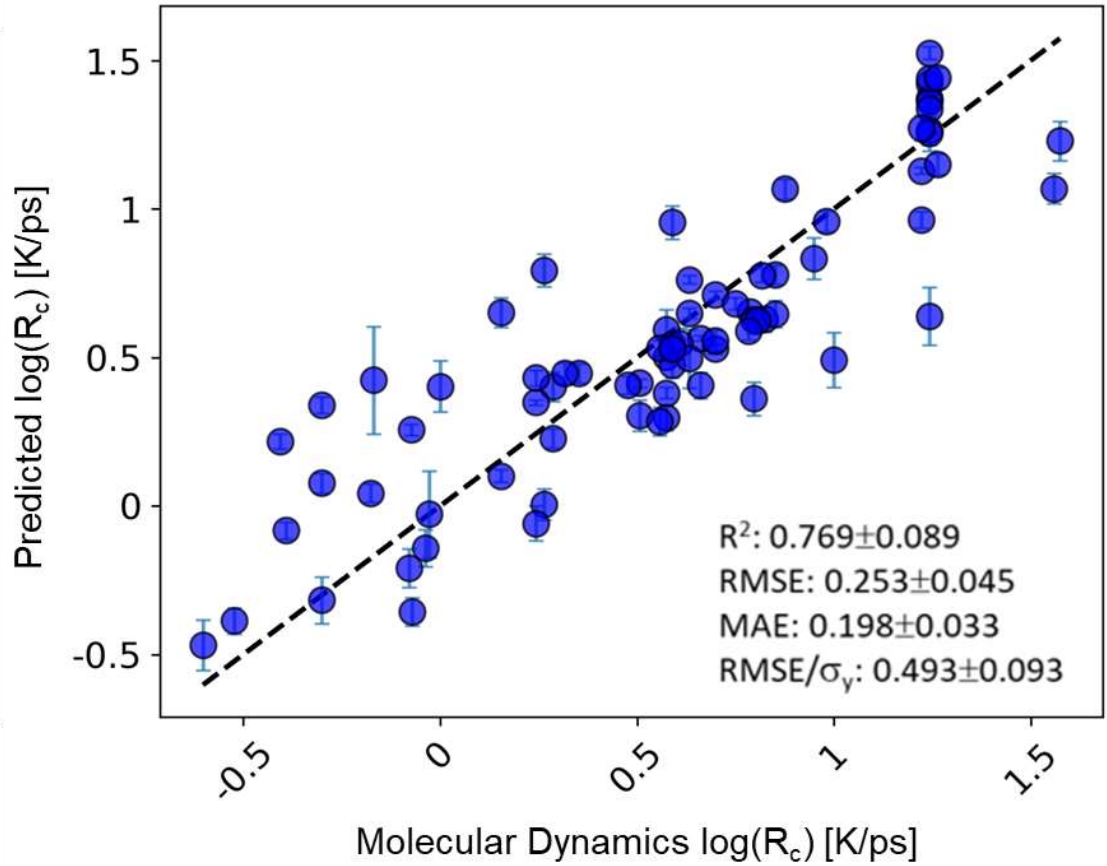

Figure 3. 5-Fold cross validation predictions of the model using the GFA feature set. Error bars show the standard deviation in predictions over 20 cross validation runs. Error metrics are averages of individual statistics from each cross validation run.

An addition to discussion of model performance we would also like to address several limitations with this model. The first is that it is important to note that the extremely high cooling rates accessible to MD simulations mean that the model only has access to training data on poor glass formers. Therefore, any predictions of good glass forming materials such as melt-spun glasses or bulk metallic glasses would be large extrapolations from the training data over multiples orders of magnitude in $R_C$. Due to the large extrapolation needed we expect any predictions of even moderately good glass formers would have large uncertainties associated with them. Furthermore, available interatomic potentials limit access to a majority of good glass forming systems which would be of the most interest to predict and compare to known GFA metrics. One binary system where we do have access to interatomic potentials is the Cu-Zr binary. Predicting $R_C$ for the $Cu_{50}Zr_{50}$ alloy gives an $R_C$ value of 0.19 K/ps which is equivalent to -0.71 on the log scale in Figure *3*. While this prediction in lower than all the training data range it is still about 13 orders of magnitude faster than the known $R_C$ value of 250 K/s [36]. This result is expected and highlights that the focus of results for training this model is not in making significant new predictions but in learning which features were important for model performance within the model's training domain.

We explored which features contributed the most to the predictive improvement. During model training, no feature selection was performed ahead of model training. The fitting process of the LASSO model performs an internal feature selection and produced a total of 48 features with non-zero coefficients. Looking at the magnitude of coefficients obtained during model fitting, we determined that a small number of features played a dominant role in the performance of the model. The magnitude of coefficients for the ten most important features is shown in *Table 2*. Note that all features have been normalized to have a minimum value of 0 and a maximum value of 1 in order to directly compare coefficients. Within 10 features, the coefficient drops by almost an order of magnitude. As a quick test for significance of features, a model was generated using only these top 10 features and it showed a RMSE of $0.25 \pm 0.04$ which is essentially identical performance to the full model shown in Figure 3. Six out of the top ten features came from the GFA features, demonstrating that the additional GFA features are critical to the improved model performance.

Table 2. Magnitudes of LASSO coefficients for the top 5 features in the second model that used both elemental features as well as the calculated GFA features are tabulated.

| Feature Name | LASSO Coefficient | Feature Class |
|---|---|---|
| enthalpy of crystallization | 2.22 | GFA |
| icosahedral-like (ICO-like) fraction (at 100K) | 1.08 | GFA |
| mendeleev number (Average) | 0.93 | ELEM |
| glass transition temperature ($T_g$) | 0.44 | GFA |
| diffusivity (at 1.2 $T_g$ [D]) | 0.44 | GFA |
| boiling temperature (average) | 0.33 | ELEM |
| APD (at 100K) | 0.31 | GFA |
| BCC crystal volume per atom (average) | 0.31 | ELEM |
| icosahedral-like (ICO-like) fraction (at 1.2 $T_g$) | 0.29 | GFA |
| number of unfilled valence orbitals (average) | 0.28 | ELEM |

Figure 4 shows the four features with highest coefficients plotted against $R_C$. Qualitative differences between the effectiveness of the four features can be readily seen. First, the $\Delta H_{crystal}$ feature shows a significantly better relationship than the other three features, with an RMSE ($R^2$) of 0.36 (0.52). The Icosahedral-like Fraction also shows some significant correlation, with an RMSE ($R^2$) of 0.46 (0.22), but the Mendeleev Number and Tg show no simple linear correlation.

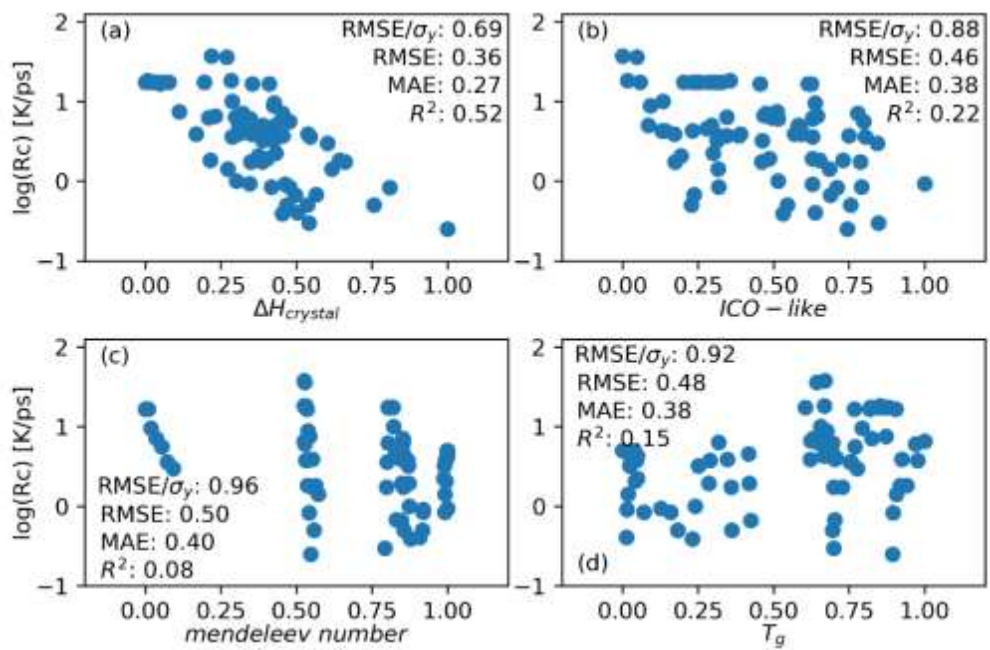

Figure 4. The top four features are plotted individually against the critical cooling rate. (a) Enthalpy of crystallization, (b) Icosahedral-like Fraction at 100K, (c) Mendeleev Number (average), (d) $T_{rg}$.

The two GFA features with the largest coefficients, the enthalpy of crystallization and the ICO-like fraction, are particularly interesting because they are quite easy to obtain computationally. Once a rapidly quenched structure is obtained, the ICO-like fraction feature can be directly calculated with minimal computational cost. The enthalpy of crystallization feature requires 3 extra calculations. However, each of these three calculations is a static energy minimization and not a computationally intensive molecular dynamics run. The accessibility of these two features means that both are practically obtainable via ab initio quantum mechanical methods and are not limited to simulations using interatomic potentials.

The enthalpy of crystallization is an estimate of the driving force for crystallization. Previous studies estimated this driving force using empirical and semi-empirical models such as those by Miedema [25,26]. This work demonstrates how a computational analog to this driving force, directly comparing simulated enthalpies of rapidly quenched amorphous structure and a reference crystal, can capture key thermodynamic information about glass forming ability from a fairly simple and fast simulation.

The ICO-like fraction has been studied experimentally and with ab initio calculations and has recently been effectively correlated with trends in glass formation ability within the Al-Sm-X, Cu-Zr, Cu-Zr-Nb, Ce-Ga-Cu, and Ni-Nb systems [20,37–39]. These studies have typically been focused on single systems and correlating GFA trends within the system with this local structure feature. This work demonstrates how the ICO-like fraction gives structural information that correlates with trends in GFA between alloy systems and not only GFA within select systems.

4. Conclusions:

Using a simulated database of Rc, we demonstrated how computationally generated features inspired by previous GFA research can be used to improve a model's ability to predict Rc of an alloy. It was found that enthalpy of crystallization and ICO-like fraction contributed most to the improved performance of the model. Both features can be practically extracted from ab initio quantum mechanical simulations, allowing them to be applied to a wide range of materials. We believe that using these simulated features in future models for GFA could significantly increase their accuracy while allowing for readily obtainable input features. We are pursuing such studies now to build on the present work.

5. Data and Codes:

Raw data for all LAMMPS simulations, complete MAST-ML inputs and outputs for machine learning runs, and data for each figure shown in this paper can be found on FigShare (https://doi.org/10.6084/m9.figshare.13202912.v1).

The most up-to-date version of MAST-ML can be found on GitHub (https://github.com/uw-cmg/MAST-ML).

6. Acknowledgements:

The authors gratefully acknowledge support from NSF DMREF award number DMR-1332851. This work used the Extreme Science and Engineering Discovery Environment (XSEDE), which is supported by National Science Foundation grant number ACI-1548562. This work used the Extreme Science and Engineering Discovery Environment (XSEDE) Stampede through allocation TG-DMR090023. This research was also performed using the compute resources and assistance of the UW-Madison Center For High Throughput Computing (CHTC) in the Department of Computer Sciences. The CHTC is supported by UW-Madison, the Advanced Computing Initiative, the Wisconsin Alumni Research Foundation, the Wisconsin Institutes for Discovery, and the National Science Foundation, and is an active member of the Open

Science Grid, which is supported by the National Science Foundation and the U.S. Department of Energy's Office of Science.